\documentclass[aps,prl,twocolumn,superscriptaddress,
,floatfix]{revtex4}
\usepackage{graphicx}
\usepackage{amsmath,amssymb}

\begin{document}

\title{Coherence and degree of time-bin entanglement from quantum dots}

\author{Tobias Huber}
\affiliation{Institut f\"ur Experimentalphysik, Universit\"at Innsbruck, Technikerstra{\ss}e 25, 6020 Innsbruck, Austria}
\author{Laurin Ostermann}
\affiliation{Institut f\"ur Theoretische Physik, Universit\"at Innsbruck, Technikerstra{\ss}e 21/3, A-6020 Innsbruck, Austria}
\author{Maximilian Prilm\"uller}
\affiliation{Institut f\"ur Experimentalphysik, Universit\"at Innsbruck, Technikerstra{\ss}e 25, 6020 Innsbruck, Austria}
\author{Glenn S. Solomon}
\affiliation{Joint Quantum Institute, National Institute of Standards and Technology  \& University of Maryland, Gaithersburg, MD 20849, USA}
\author{Helmut Ritsch}
\affiliation{Institut f\"ur Theoretische Physik, Universit\"at Innsbruck, Technikerstra{\ss}e 21/3, A-6020 Innsbruck, Austria}
\author{Gregor Weihs}
\affiliation{Institut f\"ur Experimentalphysik, Universit\"at Innsbruck, Technikerstra{\ss}e 25, 6020 Innsbruck, Austria}
\author{Ana Predojevi\'{c}}
\email[]{ana.predojevic@uibk.ac.at}
\affiliation{Institut f\"ur Experimentalphysik, Universit\"at Innsbruck, Technikerstra{\ss}e 25, 6020 Innsbruck, Austria}

\begin{abstract}
We report on the generation of time-bin entangled photon pairs from a semiconductor quantum dot via pulsed resonant biexciton generation. Based on theoretical modeling we optimized the duration of the excitation pulse to minimize the laser-induced dephasing and increase the biexciton-to-background single exciton occupation probability. This results in a high degree of entanglement with a concurrence of up to 0.78(6) and a 0.88(3) overlap with a maximally entangled state. Theoretical simulations also indicate a power dependent nature of the dephasing during the laser excitation that limits the coherence of the excitation process.
\end{abstract}


\maketitle

Single semiconductor quantum dots, due to their discrete energy structure, constitute an antibunched single photon source at a well defined frequency and with inherently sub-Poissonian statistics \cite{gauss}. They generate single photons through a recombination of an exciton, a quasi particle formed by a Coulomb-bound electron from the conduction band and a hole from the valence band. In a more refined operation mode employing biexcitons, the Coulomb-bound four-carrier states containing two electrons and two holes, quantum dots can provide pairs of photons emitted in a fast cascade very similar to the original atomic cascade experiment by Aspect \textit{et al.} \cite{aspect1982experimental}. It has been demonstrated that in the absence of the fine structure splitting of the bright exciton levels, such a cascade exhibits polarization entanglement \cite{pol1,pol2,pol3,pol4,pol5,pol6}. Entanglement of photons is a fundamental resource for long distance quantum communications \cite{Briegel, Ekert}, where it forms the central part of various quantum communication protocols like teleportation \cite{bennett} and entanglement swapping  \cite{Zhukowski}. In addition, it is an essential element of linear optical quantum computing \cite{klm}.

The ability to achieve entanglement of photons from a quantum dot is not limited to polarization.  Recently, it has been shown that the biexciton-exciton cascade can also be entangled in its emission time (time-bin) \cite{time-bin}.  This type of entanglement (encoding) is important for optical-fibre based quantum communication \cite{Honjo} due to the fact that polarization entanglement can suffer from degradation in an optical fibre outside laboratory conditions \cite{Brodsky}. In addition, a method to perform linear optical quantum computing with photons entangled in time-bin has been demonstrated recently \cite{barbieri}. Apart from the obvious goal to generate entangled photon pairs, there are further reasons for investigating the time-bin entanglement in photons emerging from a quantum dot. Namely, this method of entanglement calls for a coherent excitation and therefore is an excellent tool for investigating the coherence properties of a quantum dot system. It is precisely resonant excitation (especially two photon-resonant excitation of the biexciton \cite{twophoton, Muller2014}) combined with a quantum dot system that exhibits a high degree of coherence, that is a sine qua non for optimal use of quantum dot photons in quantum information processing.

Here, we report on an unprecedented degree of time-bin entanglement from a single quantum dot. The requirements to generate this type of entanglement include the suppression of the single exciton probability amplitude in the excitation pulse and the lowest possible degree of dephasing caused by the laser excitation. These conditions constitute contradictory demands on the excitation pulse-length and its intensity. We include a study of these limitations from an experimental and a theoretical point of view and we indicate key parameters required in order to achieve a high degree of time-bin entanglement. We indicate an optimized operation regime for the system under consideration and provide guidelines on how to extend this study to other similar systems (see supplementary material). Beyond the generation of time-bin entanglement our study also shows a generalized method to achieve a very high photon pair generation probability from quantum dots.

\begin{figure}[t]
\includegraphics[width=0.73\linewidth]{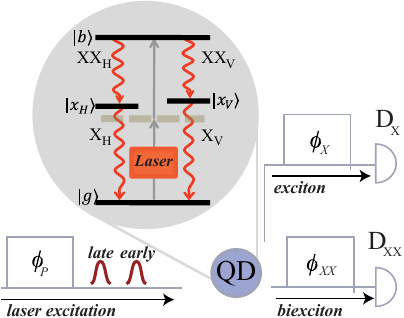}
\caption{(Color online) Schematic of time-bin entanglement. The quantum dot system (QD) is excited by two consecutive pulses obtained from an unbalanced Michelson interferometer shown on the left. The relative phase between these pulses is $\phi_P$. The state analysis is performed using additional two interferometers, one for the exciton and the other for the biexciton photons. These two interferometers have their respective phases, $\phi_X$ and $\phi_{XX}$. The photons are detected upon leaving the analysis interferometers using detectors $D_X$ and $D_{XX}$. The level scheme depicts the quantum dot excited resonantly from the ground state $|g\rangle$ to the biexciton state $|b\rangle$ using a two-photon excitation. A pulsed laser populates the biexciton via a virtual level (dashed grey line). The system decays emitting a biexciton-exciton photon cascade ($XX_V$ and $X_V$ or $XX_H$ and $X_H$). }
\label{scheme}
\end{figure}

{\em Time-bin entanglement.} This technique encodes quantum states in a superposition of the system's excitation within two distinct time-bins \textit{early} and \textit{late}. Time-bin entanglement is generated in a very similar manner for both parametric down-conversion \cite{brendel} and atom-like systems \cite{time-bin} and in its simplest form it relies on post-selection in order to be measured. Such a scheme is depicted in Fig.~\ref{scheme}. The system is addressed by two excitation pulses, denoted the \textit{early} and the \textit{late} pulse. These are derived from an unbalanced interferometer, the so-called pump interferometer. If the system is driven with a very low probability to be excited, on average only one of these two pulses will actually create a photon pair. In other words, the system is placed in a superposition of being excited by the early or by the late pulse. The relative phase, $\phi_{P}$, between the pulses determines the phase of the entangled state. This phase is written onto the quantum dot system using a coherent resonant excitation from the ground to the biexciton state. For comparison, note that in the process of parametric down-conversion the phase stability of the laser combined with the phase matching process in the extended medium ensures a constant phase relation between subsequent photon pairs. We will discuss the resonant excitation in more detail in the section on biexciton generation below.

The time-bin entangled state reads
\begin{equation}
|\Phi\rangle=\frac{1}{\sqrt2}(|early\rangle_{XX}|early\rangle_{X}+e^{i\phi_{P}}|late\rangle_{XX}|late\rangle_{X}),
\label{state}
\end{equation}
where $\phi_{P}$ is the phase of the pump interferometer generating the two excitation pulses and $|early\rangle$ ($|late\rangle$) denote photons generated in an \textit{early} (\textit{late}) time-bin. The subscripts  $XX$ and $X$ identify biexciton and exciton photons, respectively.  The analysis of the generated state is performed by two additional unbalanced interferometers, one for the exciton and one for the biexciton photons. The entangled state is observed in post-selection \cite{brendel}. 

The schematics of the three interferometers that we need to perform the experiment is presented in Fig.~\ref{scheme}. In order to control the relative phase between different interferometers we built them all within a single bulk interferometer \cite{time-bin}. An alternative approach to this would be to stabilize distinct interferometers using a narrow-band phase stable laser \cite{brendel}. The quantum dot sample was held at a temperature of 4.8~K. It contained low density self-assembled InAs quantum dots embedded in a planar micro-cavity, increasing the vertical collection of photons. The excitation light was derived from a tunable 82~MHz repetition rate Ti:Sapphire pulsed laser. The length of the laser pulses (4~ps to 12~ps) was varied by means of a pulse-stretcher. The laser wavelength was 918.7~nm, which is half way between biexciton and exciton emission (Fig.~\ref{scheme}). 

\begin{figure}[t]
\includegraphics[width=1\linewidth]{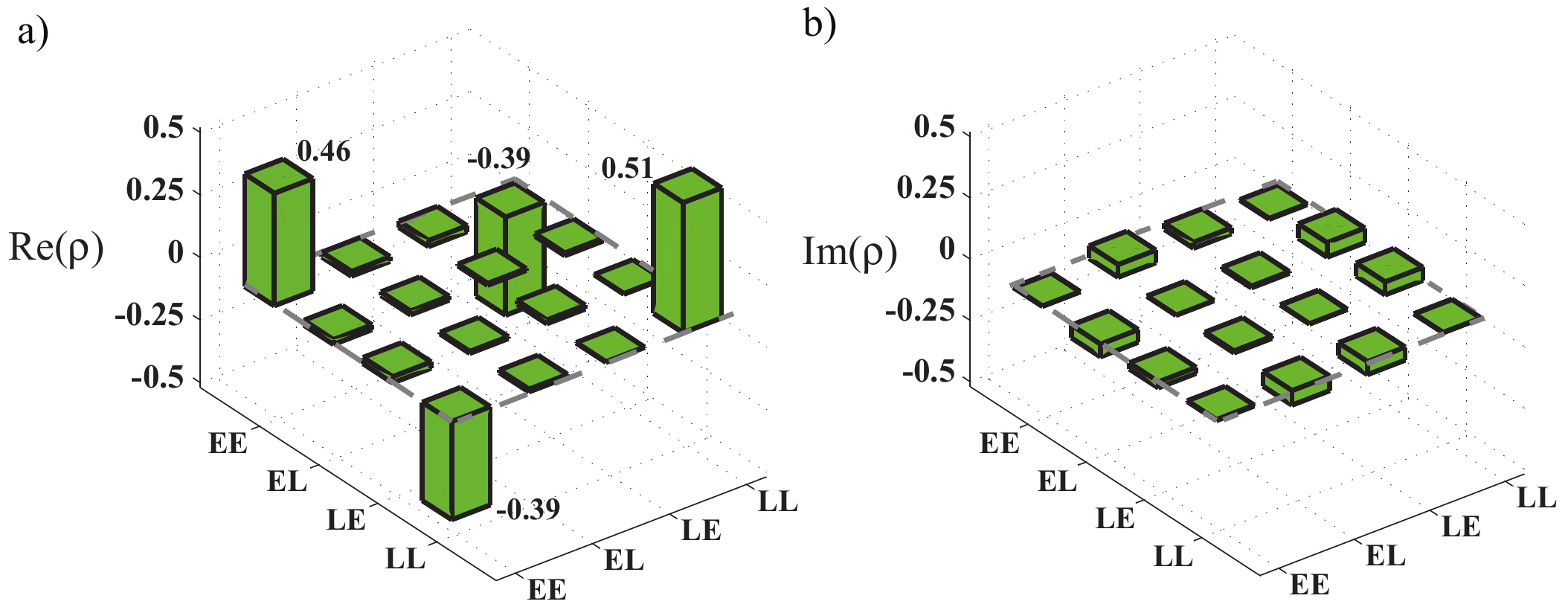}
\caption{(Color online) An example of (a) real and (b) imaginary part of the reconstructed density matrix. Measurements used to obtain this density matrix were performed using 12~ps excitation pulses while the emission probability was kept at $6 \, \%$.}
\label{matrix}
\end{figure}

We measured the degree of time-bin entanglement for three different pulse lengths: 4~ps, 9~ps, and 12~ps. To characterize the entanglement we performed state tomography where 16 projective measurements are made in three orthogonal bases (one time basis and two energy bases) \cite{james, takesue}. For a 12~ps long pulse the fidelity of the reconstructed two-photon density matrix with the maximally entangled state was found to be $F=0.88(3)$ while the concurrence was $C=0.78(6).$ The reconstructed density matrix for this case is shown in Fig.~\ref{matrix}. The values for fidelity, concurrence, and coherence of the reconstructed density matrix for the three applied pulse lengths are given in Table ~\ref{ent}.

\begin{table}[h]
\caption{Values for the fidelity with the maximally entangled Bell state, concurrence, and coherence (maximal off-diagonal element of the density matrix)}
\begin{ruledtabular}
\begin{tabular}{ccrr@{}r@{\quad}}
\multicolumn{1}{c}{pulse~length\,(ps)}& \multicolumn{1}{c}{concurrence} &
\multicolumn{1}{c}{fidelity} & \multicolumn{1}{c}{coherence}
 \\ \hline
4 & $0.56(7)$  &  $0.78(3)$ & $-0.28(3)$  \\
9 & $0.71(6)$  &  $0.83(3)$ & $-0.36(3)$  \\
12 & $0.78(6)$ & $0.88(3)$ & $-0.39(3)$ 
\end{tabular}
\end{ruledtabular}
\label{ent}
\end{table}

To give a comparison to time-bin entanglement results obtained using parametric down-conversion we measured visibilities in three orthogonal bases. For the state generated using 12~ps pulses we find the visibilities of $94(2)\, \%$ in the classically correlated basis (time basis) and $74(5)\, \%$ and $67(5)\, \%$ in the bases that indicate entanglement (energy bases).

{\em Entanglement requirements.} There are two types of factors limiting the degree of time-bin entanglement obtainable from an atom-like system: those associated with excitation and those associated with the intrinsic system coherence. The first type includes the so-called double excitations. In our measurements we drive the system with 6$\%$ probability to be excited, therefore, it will happen in $0.06^2$ of the cases that the system is excited by both the early and the late pulse. These events cause the time-basis correlations to be less than unity and they also contribute to an incoherent background in both energy bases. The effect of double excitations can be eliminated through the use of deterministic schemes for generation of time-bin entanglement \cite{simon, hughes, axel}.

The time-basis measurements are not affected by the decoherence-induced reduction of the visibility, while the energy-bases measurements are. An intuitive picture of how the decoherence affects the time-bin entanglement is the following: the pump interferometer phase, $\phi_{P}$, is transferred onto the quantum dot by means of resonant excitation. Any incoherence in the process of resonant excitation as well as in the relation between the ground and the biexciton state will lead to an uncertainty in the phase of the biexciton amplitude, which determines the relative phase of the entangled state components. This will reduce the visibility contrast and decrease the values of entanglement measures and indicators like concurrence and fidelity. 

{\em Biexciton generation.} The central goal of the photon pair generation from the quantum dot systems is to get exactly one photon at the biexciton and one photon at the exciton frequency that are produced within a short time interval and with a well defined sum phase. Luckily, the exciton and biexciton transition frequencies are well separated, so that excitation light that is tuned between these two frequencies produces a resonant two-photon coupling between the ground and the biexciton state (Fig.~\ref{scheme}). As stated above, the creation of  time-bin entanglement requires a phase stable generation of subsequent photon pairs, which is hampered by the phase uncertainty in the biexciton generation as well as any phase instability of the pump interferometer; the latter being very small in our system. A method to characterize the coherence of the excitation process is a study of the Rabi oscillations.

To predominantly generate single pairs of photons from biexciton decay, one needs to avoid populating the single exciton state as well as the decay and re-excitation of the biexciton state within one laser pulse. This creates conflicting requirements for the excitation pulse length. Namely, short pulses suppress dephasing and decay within the pulse duration but have large bandwidth and high intensity, which enhances the off-resonant generation of single excitons and power induced phase shifts. Longer pulses make the system more vulnerable to background dephasing, decay, and multiple excitations.

{\em Theoretical model.} To determine an optimized parameter regime we use a standard Lindblad master equation $\dot \rho = i \left[ \rho, H \right] + \mathcal{L} \left( \rho \right)$ based upon an effective quantum dot Hamiltonian of the form 
\begin{equation} \begin{aligned}
H &= \frac{1}{2}  \Omega(t) \left( \left| g \right \rangle \left \langle x \right| + \left| x \right \rangle \left \langle b \right| + h.c. \right)\\
&+  \left( \Delta_x - \Delta_b \right) \left| x \right \rangle \left \langle x \right| - 2 \Delta_b \left| b \right \rangle \left \langle b \right|
\end{aligned} \end{equation}
and a Liouvillian damping operator
\begin{equation}
\mathcal{L}=\sum_i \mathcal{L}_i =  \sum_i \frac{\gamma_i}{2} \left( 2 A_i^\dagger \rho A_i - A_i A_i^\dagger \rho - \rho A_i A_i^\dagger \right),
\end{equation}
where
\begin{equation}
\hat A_1 = \left | b \right \rangle \left \langle x \right|, \quad
\hat A_2 = \left | x \right \rangle \left \langle g \right| 
\end{equation}
describe exciton and biexciton decay, respectively, and the corresponding dephasing mechanisms are
\begin{eqnarray}
\hat A_{bb} =&  \left( \left | b \right \rangle \left \langle b \right| - \left | x \right \rangle \left \langle x \right| \right)  \\
\hat A_{xx} =&  \left( \left | x \right \rangle \left \langle x \right| - \left | g \right \rangle \left \langle g \right|\right).
\end{eqnarray}
In Eq. 2, $\Delta_{x}$ is the energy difference between the virtual level of the two-photon transition and the exciton energy, while $\Delta_{b}$ is the detuning between the two-photon resonance and the energy of the laser driving the system. More details can be found in the supplementary material.

\begin{figure}[h]
\includegraphics[width=0.82\linewidth]{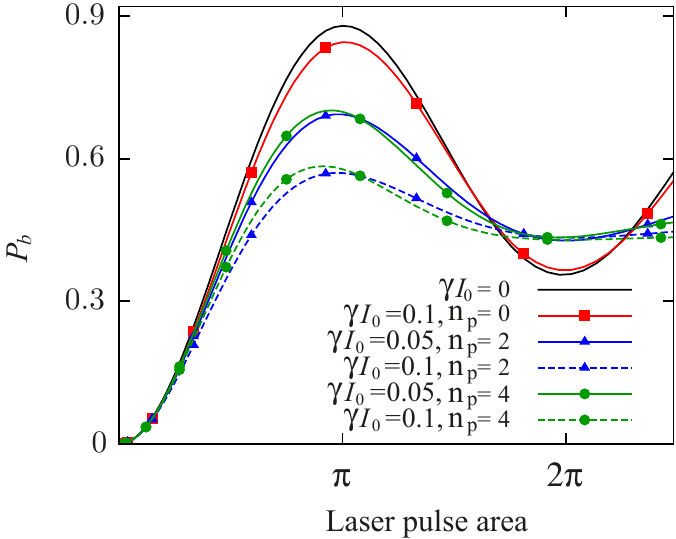}
\caption{(Color online) Biexciton emission probability, $P_b$ (theory) for different dephasing models. We observe a strong damping of the Rabi oscillations even at moderate $\gamma_{I_0}$. From this it is clear that the amplitude of the intensity-dependant dephasing rate plays a much greater role than the exponent $n_p$.}
\label{theory}
\end{figure}

We assume a Gaussian excitation laser pulse of a width $\sigma$ and a time-dependent amplitude $\Omega(t) = \Omega_0 \cdot \exp \left( - \ln(2) \left( t - t_0 \right)^2/\sigma^2 \right) $.  We calculate the emission probabilities $(P_x (t_f)$ and $P_b (t_f))$ as
\begin{equation}
P_i (t_f) = \gamma_i \int_0^{t_f} \left \langle i \right | \rho(t^\prime) \left| i \right \rangle \, \mathrm{d}t^\prime 
\end{equation}
for different pulse lengths and dephasing models. Besides a constant background dephasing rate of the freely evolving quantum dot, the presence of the laser leads to an extra and often dominant intensity-dependent dephasing rate
\begin{equation}
\gamma_{\Omega}(t)=\gamma_{I_0} \cdot \left( \Omega(t) \right)^{n_p}
\end{equation}
as detailed in \cite{muller2007}, where $\gamma_{I_0}$ is the amplitude of the intensity-dependent dephasing rate. Depending on the exponent $n_p$ of this type of dephasing, either longer or shorter pulses lead to less phase uncertainty in the generated photon pair. This phenomenon  is demonstrated in  Fig.~\ref{theory} where we compare constant and intensity dependent dephasing rates of power $n_p=0$, $2$, and $4$.

\begin{figure}[b]
\includegraphics[width=0.82\linewidth]{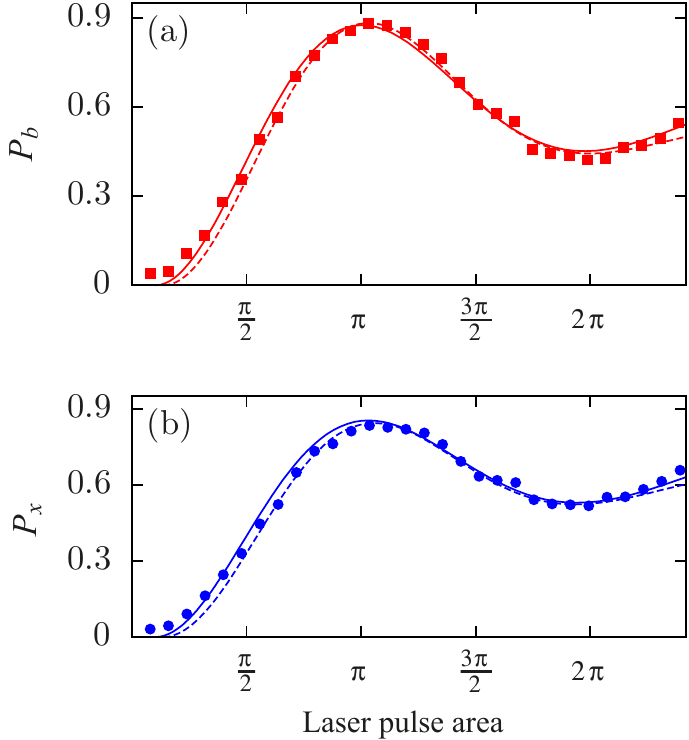}
\caption{(Color online) Emission probability for a biexciton, $P_b$, and exciton photon, $P_x$, as a function of the laser pulse area for linear (solid line) and quadratic (dashed line) intensity-dependent dephasing compared to the experimental data, respectively. The theoretical parameters, $\gamma_{I_0}$, are obtained by fitting the ratio of the first maximum and minimum of the Rabi cycle.}
\label{theory2}
\end{figure}

\begin{figure}[t]
\includegraphics[width=0.70\linewidth]{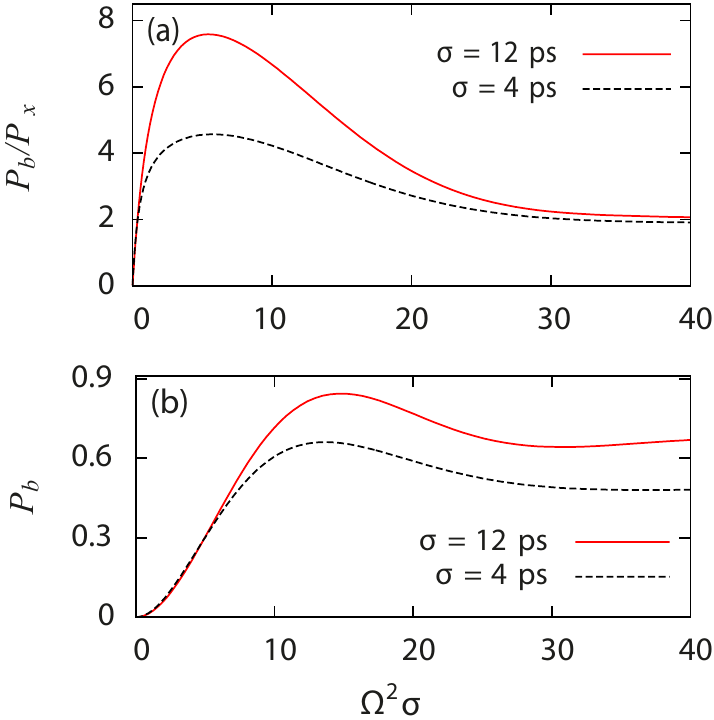}
\caption{(Color online) Relative probability for biexciton versus single exciton excitation in the quantum dot as a function of $\Omega^2\sigma$ for two different pulse lengths (upper plot). The choice of $\Omega^2\sigma$, which is proportional to the energy per pulse, as x-axis allows for easier comparison of the maxima. We see an optimum ratio of about $8$ at a still moderate excitation rate. The lower plot gives the biexciton photon emission probability as a function of $\Omega^2\sigma$.}
\label{theory3}
\end{figure}

While the constant background dephasing of the dot already yields strongly damped Rabi oscillations, both exponents $n_p=2$ and $4$ are consistent with the experimental data for dephasing rates $\gamma_{I_0} \approx 0.0349$ and $\gamma_{I_0} \approx 0.0219$, respectively, see Fig.~\ref{theory2}. Using these rates we can predict the ratio of biexcitons generated via a two-photon excitation to direct single excitons, see Fig.~\ref{theory3}. Note, that as the first is a two photon and the latter a single photon process, the second one will dominate at low powers despite of being non-resonant. As shown in Fig. \ref{theory3} better ratios are obtained at longer pulse durations due to the intensity-dependent dephasing. Note that in the lower graph of Fig.~\ref{theory2} the total exciton photon generation is depicted, which includes photons from a direct excitation of the excition as well as those generated from the decay of the biexcitons.

{\em Conclusion}. To observe the maximal degree of entanglement from single quantum dots certain  conditions need to be fulfilled. In particular, an optimized result is achieved when the quantum dot is excited in a process that exhibits a high degree of coherence. Similarly, the quantum dot  should preserve the excitation laser's phase relation. In this work we have addressed the first issue, the decoherence induced by the excitation pulse. Our theoretical study indicates that with respect to the parameters of our quantum dot system one can choose an optimized excitation pulse length.  Our measurements are consistent with the theoretical study and show a considerable degree of time-bin entanglement. The details of the calculations are given in the supplementary material. The same theoretical study can be readily extended to other quantum dot systems where it can be used to indicate the set of optimized parameters that allow for generation of high degree of time-bin entanglement. In addition, the same model will indicate the conditions needed to achieve high photon pair generation probability. Whether the latter coherence condition is fulfilled depends predominantly on the degree of interaction of the quantum dot with its semiconductor environment. The coherence can be increased relative to the lifetime of the emitted photons by the use of quantum dots embedded in micro-cavities \cite{pol2, Lermer2012} , particularly ones that are resonant to both exciton and biexciton photons \cite{pol2}. 

\begin{acknowledgments}
This work was funded by the European Research Council (project EnSeNa) and the Canadian Institute for Advanced Research through its Quantum Information Processing program. G.S.S. acknowledges partial support through the Physics Frontier Center at the Joint Quantum Institute (PFC@JQI). A. P. would like to thank the Austrian Science Fund for the support provided through project number V-375. L.O. and H.R. would also like to thank the Austrian Science Fund for support provided through SFB FoQus S4013. T. H. is receiving a DOC scholarship form the Austrian Academy of Sciences.
\end{acknowledgments}

\end{document}